%% file: smcnote.tex
\documentstyle[12pt,worldsci,epsfig]{article}
\include{feinman}

\newcommand{\ezero}{\setcounter{equation}{0}}
\newcommand{\mr}{\mathrm}

\newcommand {\oa} {\mbox{${\cal O}(\alpha)$}}

\newcommand{\bq}{\begin{equation}}
\newcommand{\eq}{\end{equation}}
\newcommand{\ba}{\begin{eqnarray}}
\newcommand{\ea}{\end{eqnarray}}
\begin{document}                                                                

\author{\ Arif~A.~Akhundov \\
\it King Fahd University of Petroleum and Minerals, \\ 
    P.O. Box 1803, Hafr Al-Batin 31991, Saudi Arabia\\
\it and \\
\it Institute of Physics, Azerbaijan Academy of Sciences, \\
\it H. Cavid ave. 33, Baku 370143, Azerbaijan}

\title{{\bf Radiative Corrections for SMC Measurements}}

\begin{center}
{\it To the memory of Prof. ENGIN ARIK (1948-2007)}
\end{center}

\maketitle                                                                      
\setlength{\baselineskip}{2.6ex}

 \begin{abstract} 
 A numerical study of QED radiative corrections for the SMC experiment
 at SPS is been performed. The semi-analytical program  POLRAD is been 
 used to get the size of the radiative corrections for the proton 
 to get the size of the radiative corrections for the measurements
 of the proton spin dependent structure functions $g_1^p(x)$ and
 $g_2^p(x)$. A brief description of the program POLRAD is given. 
 \end{abstract}
 
\section
{Introduction \label{intr}}
\ezero
 
The asymmetry of deep inelastic scattering of polarized leptons on
polarized nucleons
\bq
\l(k_1) + N(p_1) \rightarrow \l(k_2) + X(p_2),
\label{eq0}
\eq
gives important information about the spin dependent structure
functions of the nucleons $g_1(x,Q^2)$ and $g_2(x,Q^2)$ ~\cite{Rolland}.
The experimental study of the asymmetry by the Spin Muon Collaboration
(SMC) is based on the measurements of the scattering of polarized muons on polarized deuterons and protons~\cite{SMC1,SMC2,SMC3}.
In the data analysis of this
experiment the Fortran program POLRAD~\cite{POLRAD} is used 
for the calculation of the radiative corrections (RC)
from higher order QED processes~(Fig. 1) for the cross section asymmetries
\bq
  A_{\parallel} = \frac{\sigma^{\uparrow \downarrow} -
                        \sigma^{\uparrow \uparrow} }
                       {\sigma^{\uparrow \downarrow} +
                         \sigma^{\uparrow \uparrow}   }
\label{par}
\eq
 and
\bq
  A_{\perp} = \frac{\sigma^{\uparrow \rightarrow} -
                       \sigma^{\uparrow \leftarrow} }
                      {\sigma^{\uparrow \rightarrow} +
                       \sigma^{\uparrow \leftarrow}   }
\label{perp}
\eq
where $\sigma^{\uparrow\downarrow},\sigma^{\uparrow\downarrow}$
    ( $\sigma^{\uparrow\rightarrow},\sigma^{\uparrow\leftarrow}$)
are the cross sections for inclusive deep
inelastic scattering of longitudinally polarized muons off longitudinally
(transversely) polarized nucleons.
The radiative corrections are important for the higher statistics
experiments and are dependent of the methods used to extract the wanted
cross section from the data.
 
The fixed target experiments on deep inelastic lepton-nucleon scattering mainly use the energy $E'$ and the angle $\theta$ of the
scattered lepton to determine the kinematics of the events, i.e. the analysis of deep inelastic events is based  on the measurement of the familiar leptonic variables
\bq
Q_l^2 = -(k_1 - k_2)^2, \hspace{1.cm} y_l =  p_1 (k_1 - k_2)/ p_1 k_1 ,
\hspace{1.cm} x_l =  Q_l^2/(S y_l),
\label{qxyl}
\eq
with
\bq
     S = (k_1 + p_1)^2  =  2ME.
\label{s}
\eq
 Here $E$ is the energy of the incident lepton in the laboratory system,
  $\vec p_1$=0, and $M$ is the mass of the target.
 
 The methods for the calculation of the RC for deep inelastic scattering
 of leptons on unpolarized nucleons in  terms of the leptonic
 variables are well known~\cite{MoTsai,ABS}. Both methods are based
 mainly on  model-independent approaches.  The model-independent
 treatment of the RC uses the phenomenological structure functions of
 nucleons, which are defined in the Born approximation.
 A detailed comparison~\cite{Barbara} has shown that the agreement between the two methods is better than  2 \%.
 
 The model-independent radiative corrections in  terms of the
 variables $x_l, Q_l^2$
 to the cross section of deep inelastic scattering of polarized leptons
 by longitudinally polarized nucleons were calculated in~\cite{KuchShum}.
 The formulae derived in these papers were generalized in~\cite{AkuShum}
 to take into account the case of a transversely polarized target and
 the scattering on polarized spin-1 targets.

   A new development for the model-independent approach in the
 calculations of the RC for deep inelastic scattering has been arised for the
 experiments at HERA where for the physical analysis of deep inelastic events one could use not only the leptonic variables (\ref{qxyl}) but also the kinematical variables derived from the hadron measurements
\bq
 Q_h^2 = -(p_2 - p_1)^2, \hspace{1.cm} y_h =  p_1 (p_2 - p_1)/ p_1 k_1 ,
\hspace{1.cm} x_h =  Q_h^2/(S y_h),
\label{qxyh}
\eq
or some mixture of both.
 
The theoretical investigations of the radiative corrections for the experiments at HERA have been summarized in the Proceedings of the Workshop "Physics at HERA"~\cite{heraproc}, and the different programs for the  calculation of the RC have been cross-checked~\cite{hubert}.
 
 The model-independent treatment~\cite{MI} of leptonic QED corrections
 for neutral current $ep$-scattering, including
 the bremsstrahlung process
\bq
l(k_1) + N(p_1) \rightarrow l(k_2) + X(p_2) + \gamma(k),
\label{eq00}
\eq
the elastic radiative tail
\bq
l(k_1) + N(p_1) \rightarrow l(k_2) + N(p_2) + \gamma(k).
\label{eqERT}
\eq
and QED vertex corrections have been done in  terms of several
variables - leptonic, hadronic, mixed and using
the Jacquet-Blondel variables~\cite{JB}.
 
 The new formulae were implemented in the program package TERAD91~\cite{TERAD91}. The program TERAD91  can be applied to calculate the RC factor of the order {\oa} to the measured differential cross section,
$d^2 \sigma^{meas} /dxdQ^2$.
After the acceptance and smearing corrections it reduces to
\ba
   d^2 \sigma ^{meas} /dxdQ^2  =  d^2 \sigma^{Born}/dxdQ^2 ( 1 + \delta(x,Q^2)) .
\label{Born}
\ea
 
 The semi-analytical program TERAD91 has been used ~\cite{Madrid,HERMES}
 to evaluate the radiative corrections for the first
 measurement of the proton structure function $F_2(x,Q^2)$ in the
 range  $x= 10^{-2} - 10^{-4}$ and $5~GeV^2<Q^2<80~GeV^2$
 using the electron and the mixed kinematical variables ~\cite{H1F2}.
 
 The radiative correction factor $\delta(x,Q^2)$ for the inclusive cross
section of deep inelastic scattering is defined by the equation
\ba
\delta({x,Q^2})=
 \frac{d^2{\sigma}^{\mr{theor}}/{dxdQ^2}}
      {d^2{\sigma}^{\mr{Born} } /{dxdQ^2}}-1 ,
\label{delta}
\ea
where  $ d^2{\sigma}^{\mr{theor}}/{dxdQ^2} $ is the
theoretical approximation for the measured cross section,
$ d^2{\sigma}^{\mr{meas}}/{dxdQ^2} $, which contains contributions
from higher order QED and electroweak processes, and
    $ d^{2}{\sigma}^{\mr{Born}}/{dxdQ^2} $
is the Born cross section of the process
(\ref{eq0}).
 
In this note we describe the semi-analytical program POLRAD which has been
used to calculate the radiative corrections for
the measurements of the spin dependent proton structure
functions $g_1^p(x)$ and $g_2^p(x)$ by the SMC experiment.

\section
{The semi-analytical program POLRAD
\label{Polrad}}
\setcounter{equation}{0}
 
    For  $Q^2~\ll~M_z^2$ only electromagnetic  processes are
    relevant, and   the unpolarized part of the Born cross section
   of the deep inelastic scattering (\ref{eq0})
is determined by two structure functions  $F_2(x,Q^2)$ and
\bq
  2xF_1(x,Q^2)=\frac{(1+\rho^2)F_2(x,Q^2)}{1~+~R(x,Q^2)} ;
\label{F1}
\eq
\bq
  \frac{d^2\sigma_0^{unpol}}{dxdQ^2}  = \frac{4\pi \alpha^2}{Q^4 x}
\left[ 1-y ~+~\frac{y^2}{2(1+R(x,Q^2))} \right]   F_2(x,Q^2),
\label{Bornunpol}
\eq
 Here and below
\bq
    x \equiv x_l, \hspace{1.cm} y \equiv y_l, \hspace{1.cm} Q^2 \equiv Q_l^2
\label{qxy}
\eq
and
\ba
     \rho^2=\frac{Q^2}{\nu^2}=\frac{2Mx}{Ey}.
\label{ro}
\ea
 
The spin dependent part of the Born cross section has contributions from the
structure functions $g_1$ and $g_2$ as:
\bq
  \frac{d^2\sigma_0^{\uparrow\downarrow}}{dxdQ^2}
  =  \frac{8\pi \alpha^2}{Q^4 }
 y\left[~(1-\frac{y}{2}-\frac{\rho^2y^2}{4})g_1(x)~-~\frac{\rho^2y}{2}g_2(x)~\right] ,
\label{Bornlong}
\eq
 when the beam and target spins are collinear, and
\bq
  \frac{d^2\sigma_0^{\uparrow\rightarrow}}{dxdQ^2}
     = \frac{8\pi \alpha^2}{Q^4 }
 y\rho\sqrt{1-y-\frac{\rho^2y^2}{4}}\left[\frac{y}{2}g_1(x)~+~g_2(x) \right] ,
\label{Borntran}
\eq
 when the spin directions of beam and target are orthogonal.
 
 The program POLRAD contains the formulae for leptonic QED corrections
 to the cross sections (\ref{Bornunpol}),(\ref{Bornlong})
 and (\ref{Borntran}). These corrections
 include photonic bremsstrahlung from leptons, vertex corrections (see
 Fig.1, only $\gamma$-exchange ) and
 vacuum polarization of the order {\oa} with soft-photon exponentiation.
 This set of the higher order Feynman graphs is standard for the model-independent treatment of the QED RC to deep inelastic scattering~\cite{ABS} and gives the numerically largest contribution.
 
 The general formula for the radiative cross sections expressed in
 terms of  leptonic variables can be written as a sum of two parts:
 
\bq
  \frac{d^2\sigma^{\oa}}{dxdQ^2}  =
  \frac{d^2\sigma^{Born}}{dxdQ^2}~(1~+~\delta^{VR}(x,Q^2))+
 \int\int{dx_hdQ_h^2}~H(x,Q^2,x_h,Q_h^2)~\frac{d^2\sigma^{Born}}{dx_hdQ_h^2}
\label{RC}
\eq
 The first part of (\ref{RC}), proportional to $\delta^{VR}$, contains
 the contributions from vertex corrections, vacuum polarization and
 from the soft part of the real photon radiation. The second part
 accounts for the bremsstrahlung of hard photons. It depends on the
 structure functions, not only in a given $ F(x,Q^2) $ point ,
 but in the range of $(x_h,Q_h^2)$ defined by kinematics of Fig. 2a.
 
For radiative events
\bq
      x~\leq~x_h~\leq~1, \hspace{1.cm}
      M^2~\leq~M_h^2~\leq~W^2 ,
\label{xM}
\eq
 were the square of the invariant
mass of the hadronic final state $M_{h}^{2}$ can be defined as
\begin{equation}
M_{h}^{2} = M^{2} + Q_h^2  \left( \frac{1-x_{h}}{x_{h}} \right) ,
\end{equation}
  and  the invariant mass
\begin{equation}
W^{2} = M^{2} + Q^2 \left( \frac{1-x}{x} \right),
\end{equation}
 corresponds to the hadronic jet without radiation.
 
 From the experience of the studies of the RC for deep inelastic
 scattering~\cite{ABS,hubert} it is known that the large radiative
 corrections are mainly due to the emission of hard photons
 ( the second part of (\ref{RC})). For hard photon
 bremsstrahlung in the process (\ref{eq00}) and (\ref{eqERT})  the
 kinematic minimum of $Q_h^2$ is
\bq
       ({Q_h^2})_{min}~\simeq~ x^2 M^2.
\label{Qmin}
\eq
 This formula shows that it is impossible to calculate the RC without
 the knowledge of the nucleon structure functions
 in the region $Q^2 \rightarrow 0 $.
 
 The program POLRAD, instead using of the usual $M_h^2$ and $Q_h^2$,
 or $x_h$ and $Q_h^2$, (Fig. 2a) is using other invariant kinematical variables (Fig. 2b):
\bq
    R = W^2-M_h^2+Q^2-Q_h^2 ,\hspace{1.cm}
    \tau = \frac{Q_h^2-Q^2}
                           {W^2-M_h^2+Q^2-Q_h^2},
\label{Rtau}
\eq
 for the calculation of the two-dimensional integral in (\ref{RC})--
 the contribution of the inelastic radiative tail.
 The boundaries of these variables are:
\bq
  0~\leq~R~\leq ~\frac{W^2- M^2}{1~+~\tau},
\label{Rlim}
\eq
and
\bq
  \tau_{max,min} = \frac{Sy}{2M^2}(1~\pm~\sqrt{1~+~\frac{M^2x}{Sy}}).
\label{taulim}
\eq
 The point $F(x,Q^2)$ of the Fig.2a in which we want to calculate
 the RC has been transformed into the line $R~=~0$ from $\tau_{min}$
 to $\tau_{max}$ of the Fig. 2b, and the Born kinematics of the process
 (\ref{eq0}) is defined on this line. The contribution of the elastic
 radiative tail (\ref{eqERT}) is determined by the integral on the line
 $ R~=~(W^2-M^2)/(1+\tau)$.
 
  Now we briefly explain how the program POLRAD works.
 For any point $F(x,Q^2)$ of deep inelastic scattering  the POLRAD
 routine CONKIN evaluates the Born kinematics (\ref{Rtau})-(\ref{taulim}) and
 routine BORNIN  gives the Born cross sections (\ref{Bornunpol})-(\ref{Borntran}).
 DELTAS calculates the factorized part of the RC $\delta^{VR}$
 which is independent on the polarization of the beam or target.
 The two-dimensional integral in (\ref{RC}) is calculated in terms
 of the variables (\ref{Rtau}) by the routines QQT, QQINT, TAILS
 and FFU. The routines STRF and COMFST contain the parameterizations of the
 structure functions $F_{1,2}$ , $g_{1,2}$
 and also elastic and quasi-elastic form factors for the calculation
 of their contribution to the process (\ref{eqERT}). In the last two
 routines there is the possibility of putting other parameterizations of
 structure functions and form factors.
 The main routine POLRAD calculates the final results --
 the RC to the asymmetries.

\section
 {Numerical results and conclusion
\label{results}}
\setcounter{equation}{0}
 In this section we will give the results of the calculations of the RC with the help of the program POLRAD to the longitudinal $A_{\parallel}^p$ and transverse  $A_{\perp}^p$ asymmetries in inclusive deep inelastic scattering of polarized muons off polarized protons.
 
 In order to calculate the RC one has to use the realistic parameterizations
 of the structure functions $F_{1,2}$ and $g_{1,2}$ over the full range
 of $x$ and $Q^2$. Several different parameterizations of the
 unpolarized structure functions, $F_2^p(x,Q^2)$,
  including the NA37 parametrization ~\cite{NA37}, have been used.
  For the region $ Q^2~<~0.2~GeV^2 $ the parameterizations
 of $F_2(x,Q^2)$  were
 taken at $Q^2=0.2~GeV^2$ and
  multiplied by the suppression factor~\cite{Prokh}
 $ (1 - exp(-aQ^2)) $ , where  $ a=3.37~GeV^{-2}$.
 The $F_1^p(x,Q^2)$ is calculated by (\ref{F1}), where
 the values of $R(x,Q^2)$ were taken from the fit of the SLAC data
 ~\cite{E140}.
 
 The longitudinal spin structure function, $g_1^p$, can be obtained
 from the asymmetry $A_1^p$ by the relation
\bq
    g_1^p(x,Q^2)=\frac{A_1^p(x,Q^2) F_2^p(x,Q^2)}
                      {2x\left[1~+~R(x,Q^2)\right]},
\label{g1}
\eq
where $A_1^p$ is the virtual photon asymmetry which can be measured
experimentally. But here we will use the model
parameterizations for $g_1^p$ from~\cite{Sch,Wol,Alt}.
 
 For the calculation of the transverse asymmetry, $A_{\perp}^p$,
  $g_2^p$ is needed. It can be estimated
 by equation (23) of ref.~\cite{Rolland} if we neglect the contribution
 of the last term of this equation.
 
 Figures 3-9 shows the results for the  quantities
\bq
    \frac{A_{\parallel}^p}{D_{\parallel}} , \hspace{0.5cm}
    \frac{\Delta A_{\parallel}^p}{D_{\parallel}}, \hspace{1.cm}
    \frac{A_{\perp}^p}{D_{\perp}}, \hspace{0.5cm}
    \frac{\Delta A_{\perp}^p}{D_{\perp}},
\label{A^p}
\eq
where $\Delta A^p$ is the difference between the asymmetries
 with and without the radiative corrections
\bq
     \Delta A^p~=~A^p(Born+QED~RC)~-~A^p(Born).
\label{Delta}
\eq
and $D_{\parallel,\perp}$ are the depolarization factors
\bq
  D_{\parallel}=  \frac{y(2-y)}{y^2+2(1-y)(1+R)}, \hspace{1.cm}
  D_{\perp}=  \frac{ 2y\sqrt{1-y}}{y^2+2(1-y)(1+R)}.
\label{D}
\eq
The calculations presented here have been performed with the NA37
parametrization~\cite{NA37} for $F_2(x,Q^2)$.
 
From Figures 5 and 6 it is seen that for the longitudinal polarization
of the protons we get different results for $\Delta A_{\parallel}^p/
D_{\parallel}$ and we see a
strong dependence of the RC on the choice of the $g_1^p$ parametrization.
The RC are not small at low $x$ :
       $\mid~\Delta A_{\parallel}^p/D_{\parallel}~\mid~\leq~0.025$.
 At low $x$ the influence of the hard bremsstrahlung
from the processes (\ref{eq00}) and (\ref{eqERT}) on the value        
of the RC is very
big.                                                         
For the results which were obtained from POLRAD in the case of the
transverse polarization of the target we have used the $g_1^p$
parametrization from~\cite{Alt} and have calculated $g_2^p$ from
eq. 23 of~\cite{Rolland}.
The transverse asymmetry with such input for $g_{1,2}^p$
in the Born approximation looks good for the who kinematical region.
But for the QED corrected
  asymmetry $A_{\perp}^p/
D_{\perp}$ (Fig. 8) and for absolute correction
 $\Delta A_{\perp}^p/ D_{\perp} $ (Fig. 9)
 we get results which cannot be understood.
 
It is worth mentioning that the treatment of the radiative corrections should be cross–check with the results of other programs.
From this point of view the comparison of the results of the semi-analytical program POLRAD with the results of other existing program is needed.  For instance, with the Monte Carlo event generator HERACLES~\cite{HERACLES} where the independent calculations have been implemented~\cite{Zeuthen}.
 
\section*{Acknowledgements}
I am indebted to M.~Velasco for the help in the calculations and in 
preparing the figures. I would like to thank R.~Voss and R.~Windmolders
for support of this work. I thank L.~Klostermann and S.~Rock for the helpful discussion and A. Shiekh for the comments.
 
It is a great pleasure to thank Prof. R\"udiger Voss for the kind invitation to visit CERN in 1994 and the SMC Collaboration for the good opportunities to work at CERN.


\newpage

\setlength{\textwidth}{168mm}
\setlength{\textheight}{240mm}
\setlength{\oddsidemargin}{-0.3cm}

\vspace*{2.0cm}
\begin{minipage}[t]{7.8cm}{
\begin{center}
   Born
\vspace*{0.5cm}
\begin{Feynman}{60,60}{0,0}{0.8}
%
\put(00,60){\fermiondrr}
\put(-11,62){$l^-(\vec k_1,m)$}  
\put( 50,62){$l^-(\vec k_2,m)$}
\put( 50,-06){$X(\vec p_2,M_h)$}
\put(-09,-05){$N \, (\vec p_1,M)$}
\put(34,30){$\gamma, Z$}
\put(60,60){\fermionurr}
\put(30.5,15){\photonup}
\put(30,15){\circle*{5}}
\put(30,17){\line(2,-1){30}}
\put(30,13){\line(2,-1){28}}
\put(30,45){\circle*{1.5}}
\put(30,15){\fermionurr}
\put(30,15){\fermiondrr}
\end{Feynman}
\end{center}
}\end{minipage}
\begin{minipage}[t]{7.8cm} {
\begin{center}
   vertex correction
\vspace*{0.5cm}
\begin{Feynman}{60,60}{0,0}{0.8}
%
\put(34,30){$\gamma, Z$}
\put(29,57){$\gamma   $}
\put(00,60){\vector(2,-1){13.5}}
\put(11,54.5){\line(2,-1){19.}}
%
\put(45,52.5){\line(2,1){15}}
\put(30,45){\vector(2,1){19.}}
\put(14.5,52.4){\photonright}
\put(45.0,52.4){\circle*{1.5}}
\put(15.0,52.4){\circle*{1.5}}
\put(30,45){\circle*{1.5}}
\put(30.5,15){\photonup}
\put(30,15){\circle*{5}}
\put(30,17){\line(2,-1){30}}
\put(30,13){\line(2,-1){28}}
\put(30,15){\fermionurr}
\put(30,15){\fermiondrr}
\end{Feynman}
\end{center}
}\end{minipage}

\vspace*{2.cm}
 
\begin{minipage}[tbh]{7.8cm}{
\begin{center}
  initial state radiation
\vspace*{0.5cm}
\begin{Feynman}{60,60}{0,0}{0.8}
%
\put(00,60){\fermiondrr}
\put(60,60){\fermionurr}
\put(30.5,15){\photonup}
\put(30,15){\circle*{5}}
\put(24.5,56.25){$\gamma \, (\vec k)$}
\put(34,30){$\gamma, Z$}
\put(30,17){\line(2,-1){30}}
\put(30,13){\line(2,-1){28}}
\put(30,45){\circle*{1.5}}
\put(30,15){\fermionurr}
\put(30,15){\fermiondrr}
\put(7.5,56.25){\circle*{1.5}}
\put(7.0,56.25){\photonrighthalf}
\end{Feynman}
\end{center}
}\end{minipage}
\begin{minipage}[tbh]{7.8cm} {
\begin{center}
  final state radiation
\vspace*{0.5cm}
\begin{Feynman}{60,60}{0,0}{0.8}
%
\put(00,60){\fermiondrr}
\put(60,60){\fermionurr}
\put(30.5,15){\photonup}
\put(71,56){$\gamma \, (\vec k)$}
\put(34,30){$\gamma, Z$}
\put(30,15){\circle*{5}}
\put(30,17){\line(2,-1){30}}
\put(30,13){\line(2,-1){28}}
\put(30,45){\circle*{1.5}}
\put(30,15){\fermionurr}
\put(30,15){\fermiondrr}
\put(53.0,56.25){\photonrighthalf}
\put(53.0,56.25){\circle*{1.5}}
\end{Feynman}
\end{center}
}\end{minipage}

\vspace*{2.cm}

Fig. 1. Feynman graphs contributing in the higher order to deep inelastic
        $lN$-scattering. 
 
\newpage

\begin{figure}
\epsfysize22cm
\epsffile{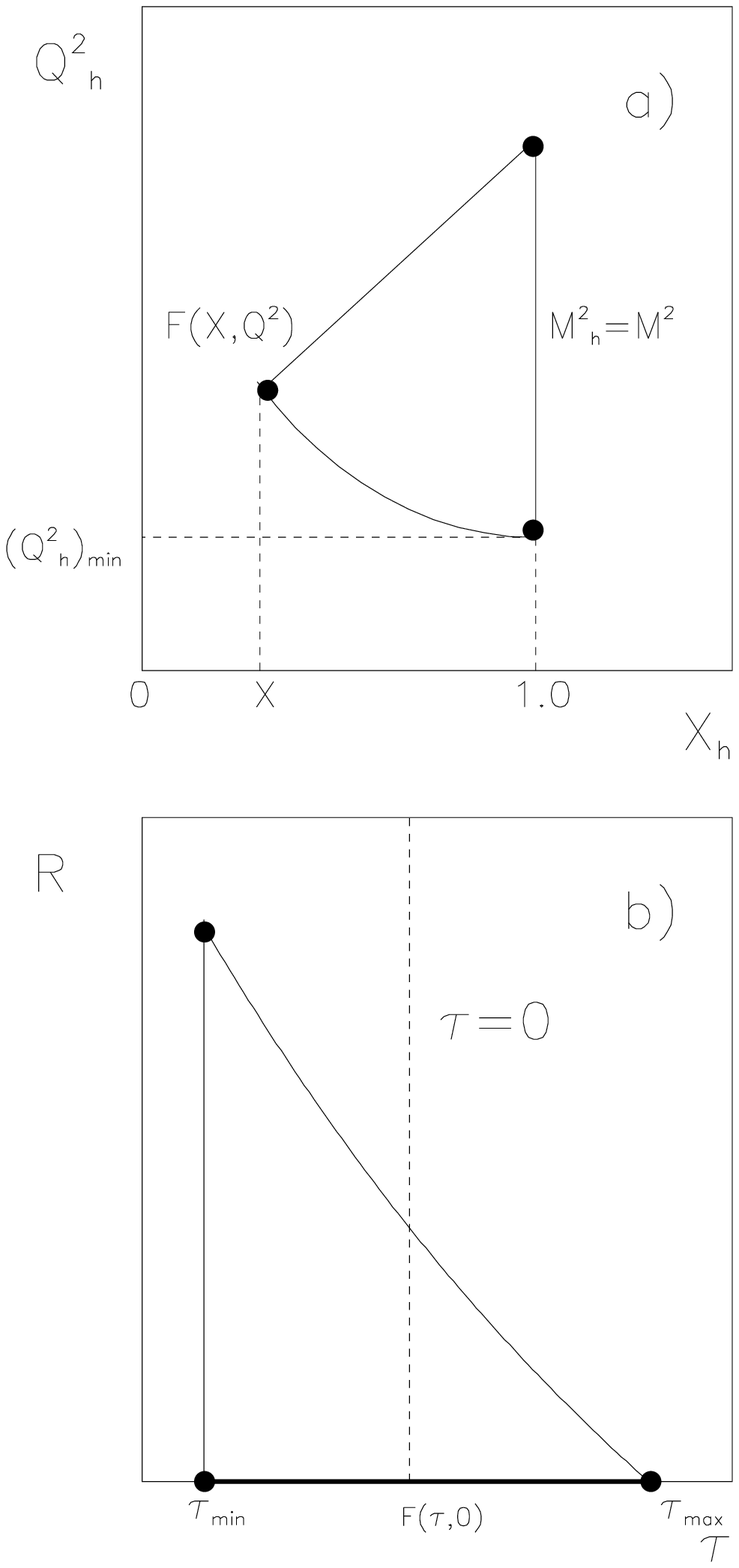}
\begin{center}
{Fig. 2. Integration region of $(x_h,Q_h^2)$ in (\ref{RC})
  and the correspondent region in the terms of $(\tau,R)$.}
\end{center}
\end{figure}

\begin{figure}
\epsfysize22cm
\epsffile{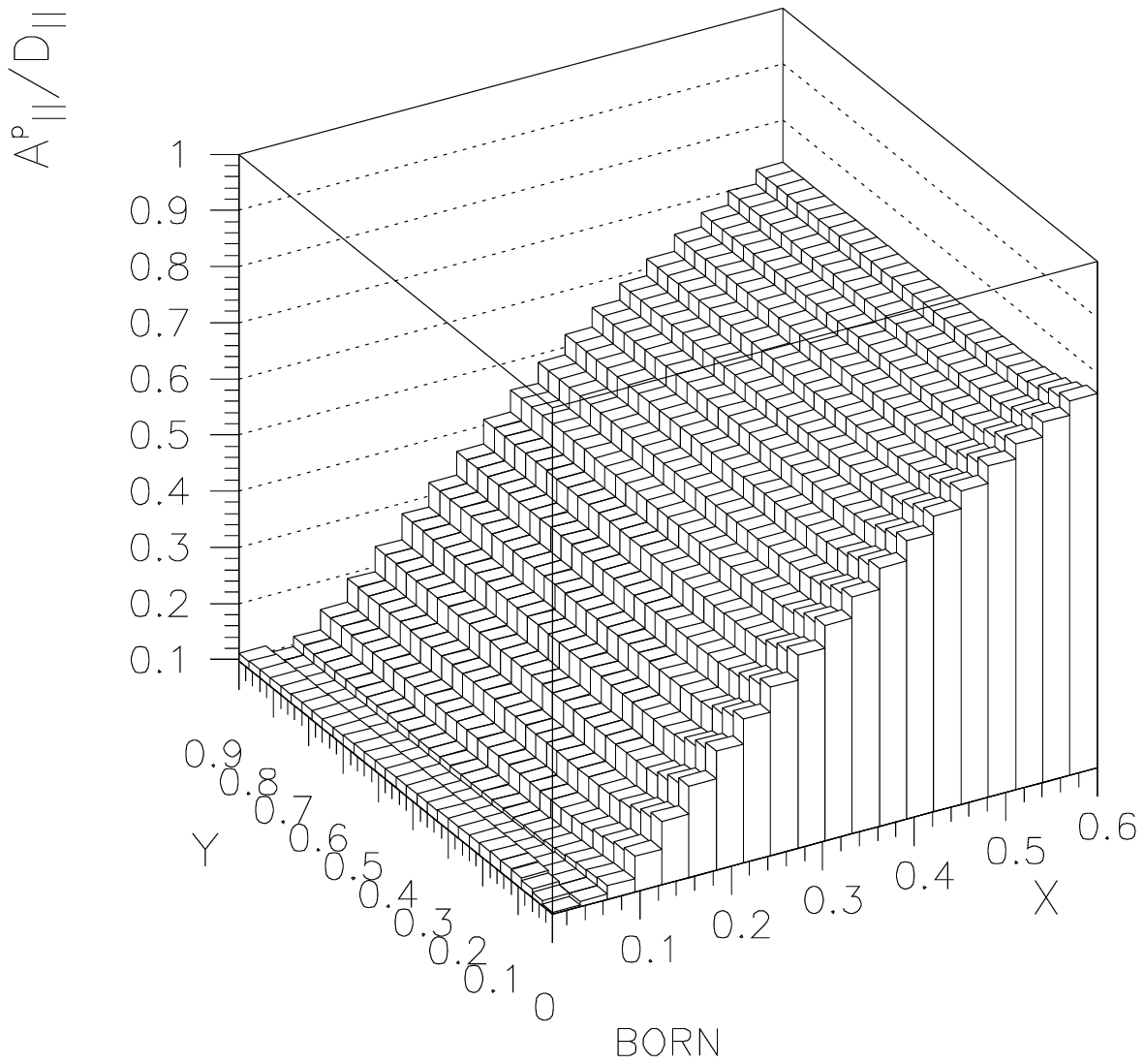}
\begin{center}
{Fig. 3. Longitudinal asymmetry $ A_{\parallel}^p/
                                             D_{\parallel}$
in the Born approximation for beam energy $ 190~GeV^2$;
$g_1^p$ from ~\cite{Wol}.}
\end{center}
\end{figure}

\begin{figure}
\epsfysize22cm
\epsffile{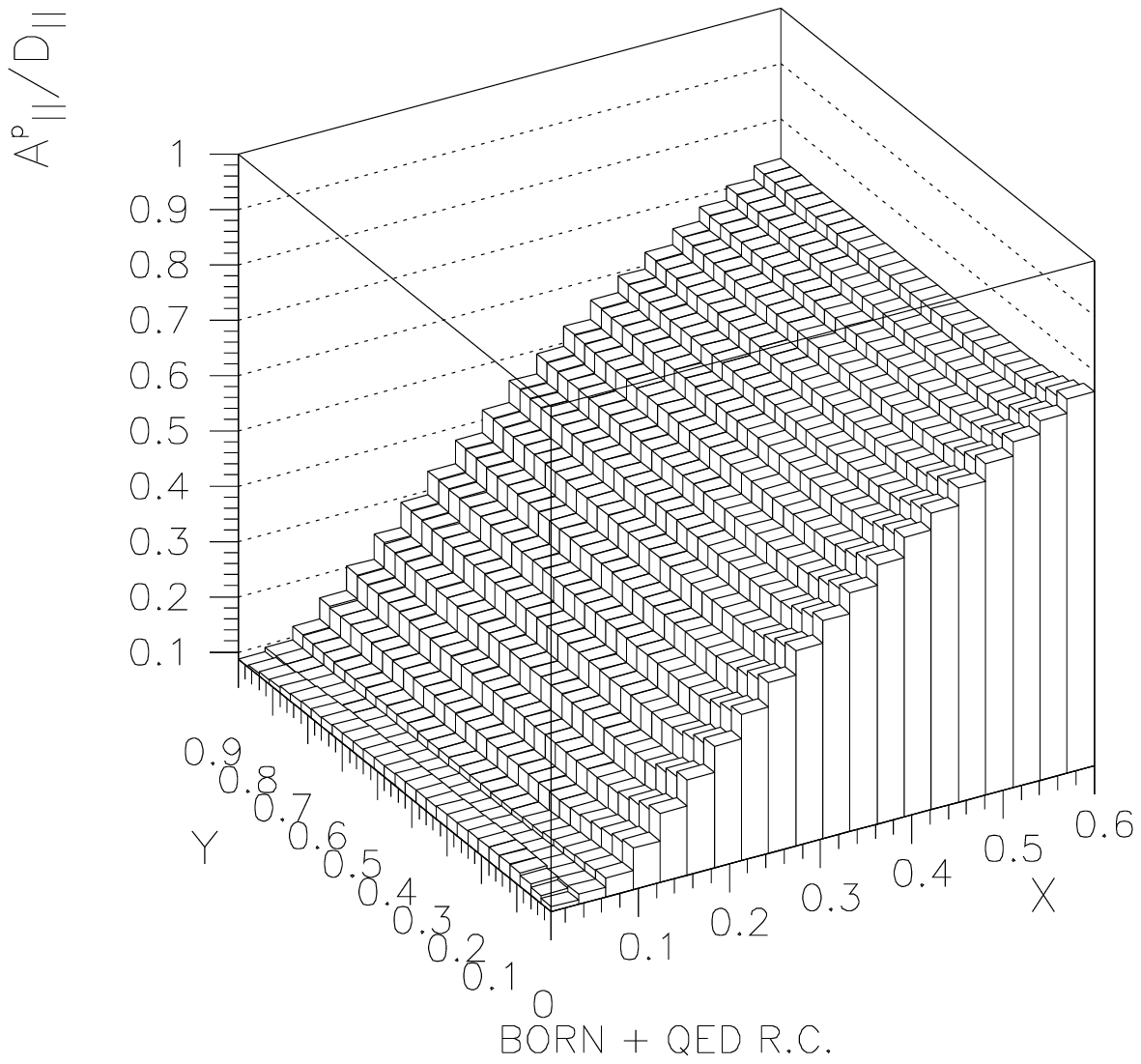}
\begin{center}
{Fig. 4.  QED corrected
               longitudinal asymmetry $  A_{\parallel}^p/
                                             D_{\parallel} $
                          for beam energy $ 190~GeV^2$;
$g_1^p$  from ~\cite{Wol}.}
\end{center}
\end{figure}

\begin{figure}
\epsfysize22cm
\epsffile{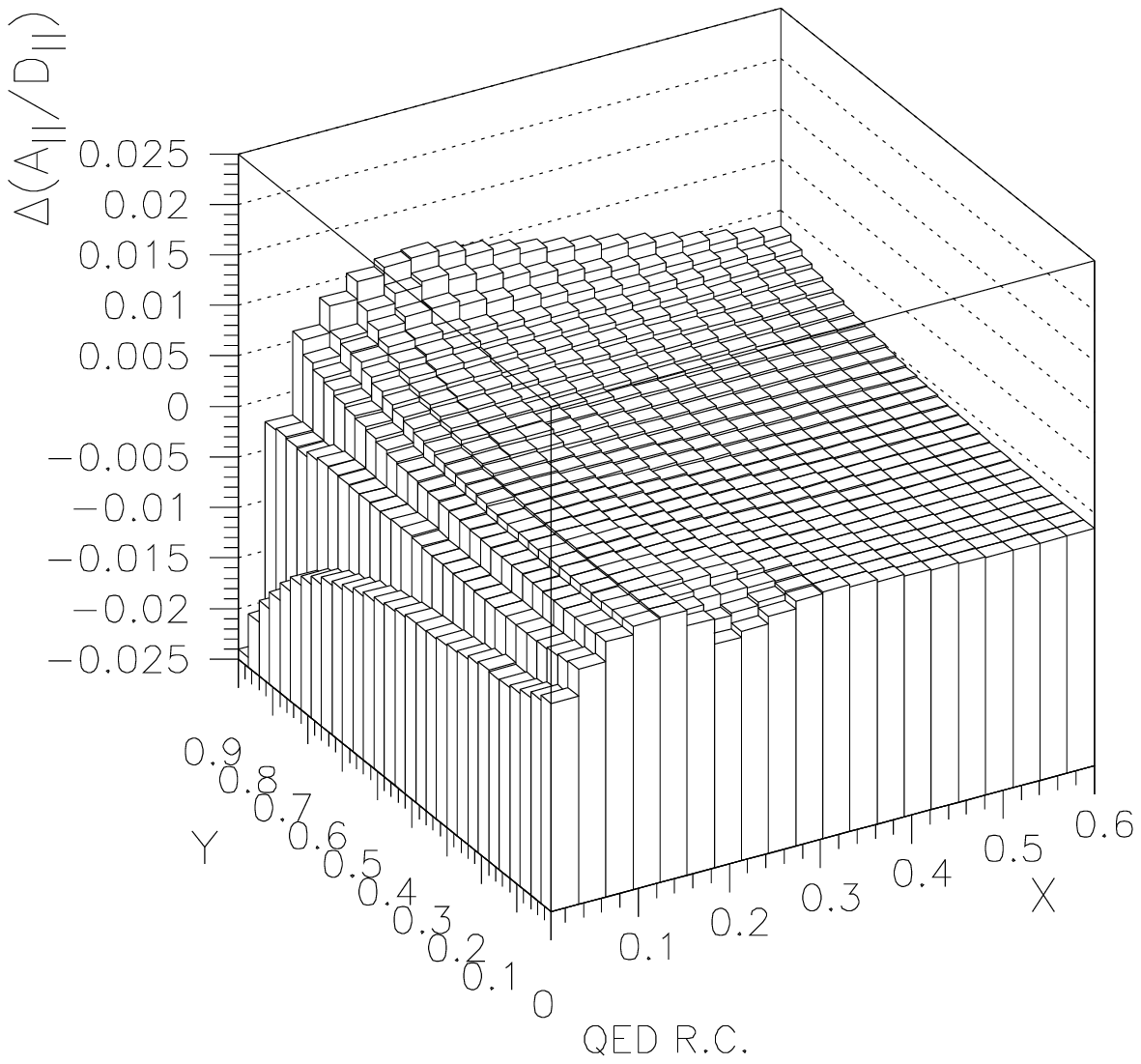}
\begin{center}
{Fig. 5.   Absolute correction  $  \Delta A_{\parallel}^p/
                                             D_{\parallel} $
                          for beam energy $ 190~GeV^2$;
$g_1^p$  from ~\cite{Wol}.}
\end{center}
\end{figure}
 
\begin{figure}
\epsfysize22cm
\epsffile{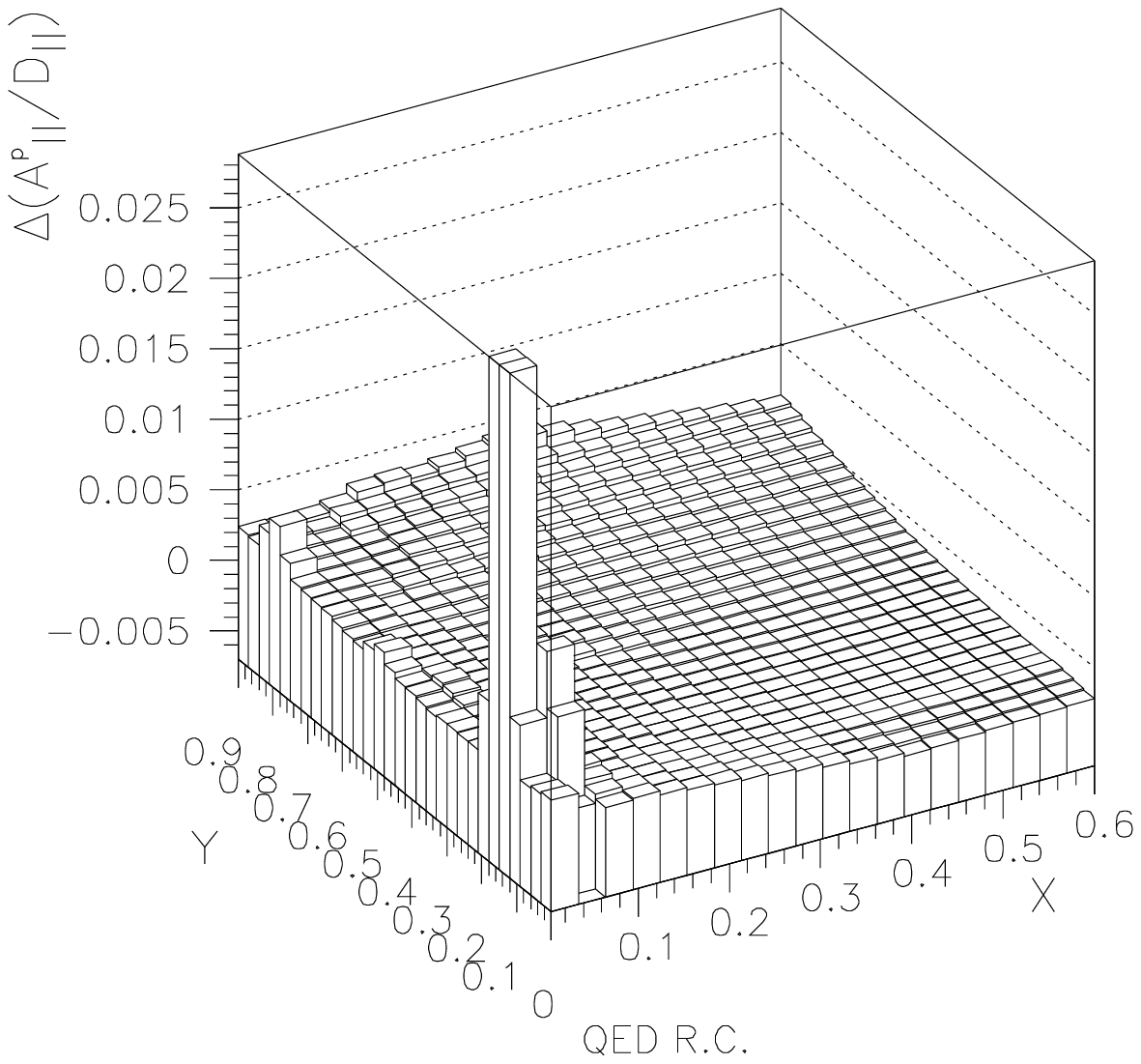}
\begin{center}
{Fig. 6.     Absolute correction  $  \Delta A_{\parallel}^p/
                                             D_{\parallel}$
                          for beam energy $ 190~GeV^2$;
$g_1^p$  from ~\cite{Sch}.}
\end{center}
\end{figure}

\begin{figure}
\epsfysize22cm
\epsffile{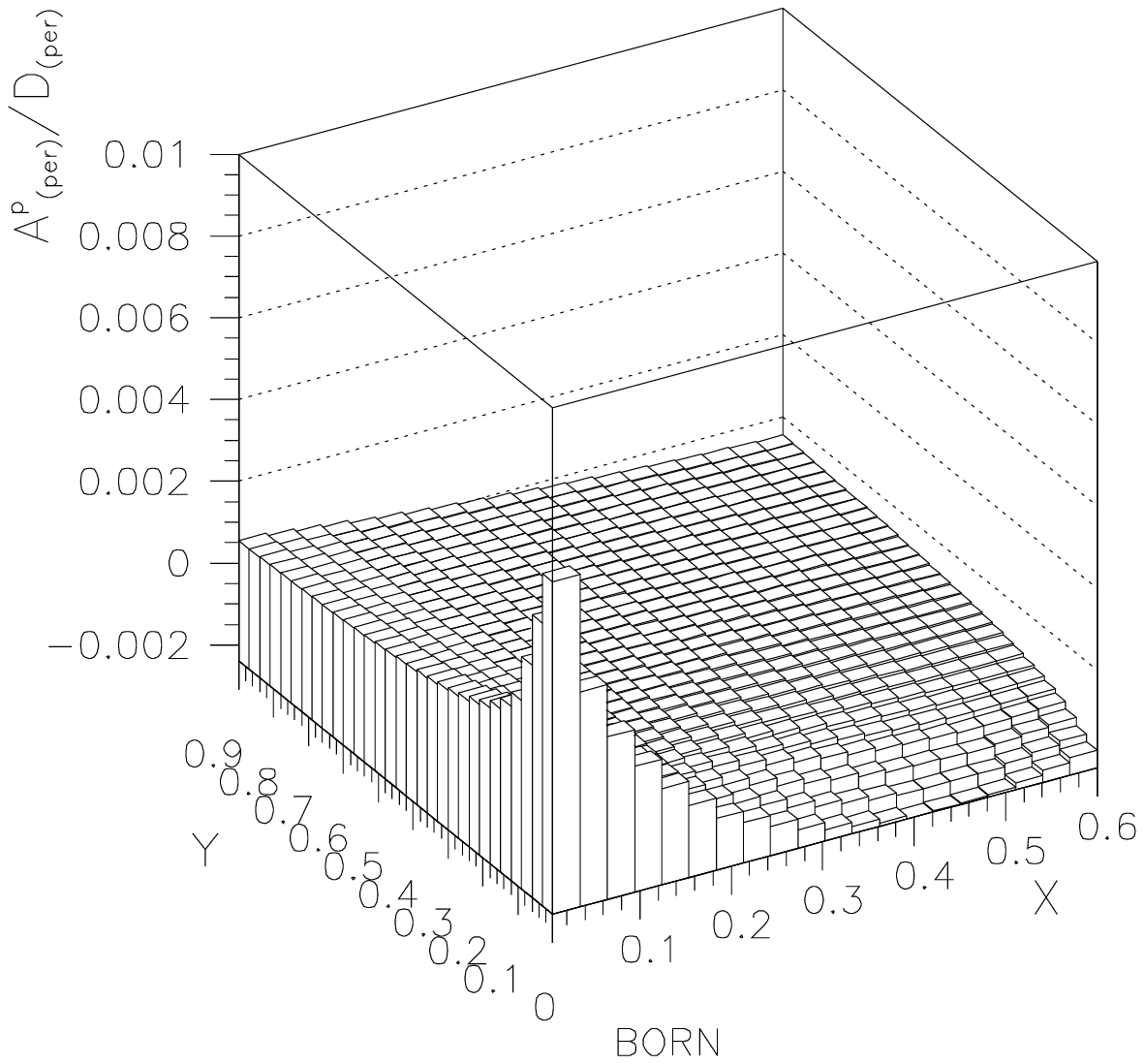}
\begin{center}
{Fig. 7.     Transverse  asymmetry $ A_{\perp}^p/
                                             D_{\perp} $
in the Born approximation for beam energy $ 100~GeV^2$;
$g_1^p$  from ~\cite{Alt}.}
\end{center}
\end{figure}
 
\begin{figure}
\epsfysize22cm
\epsffile{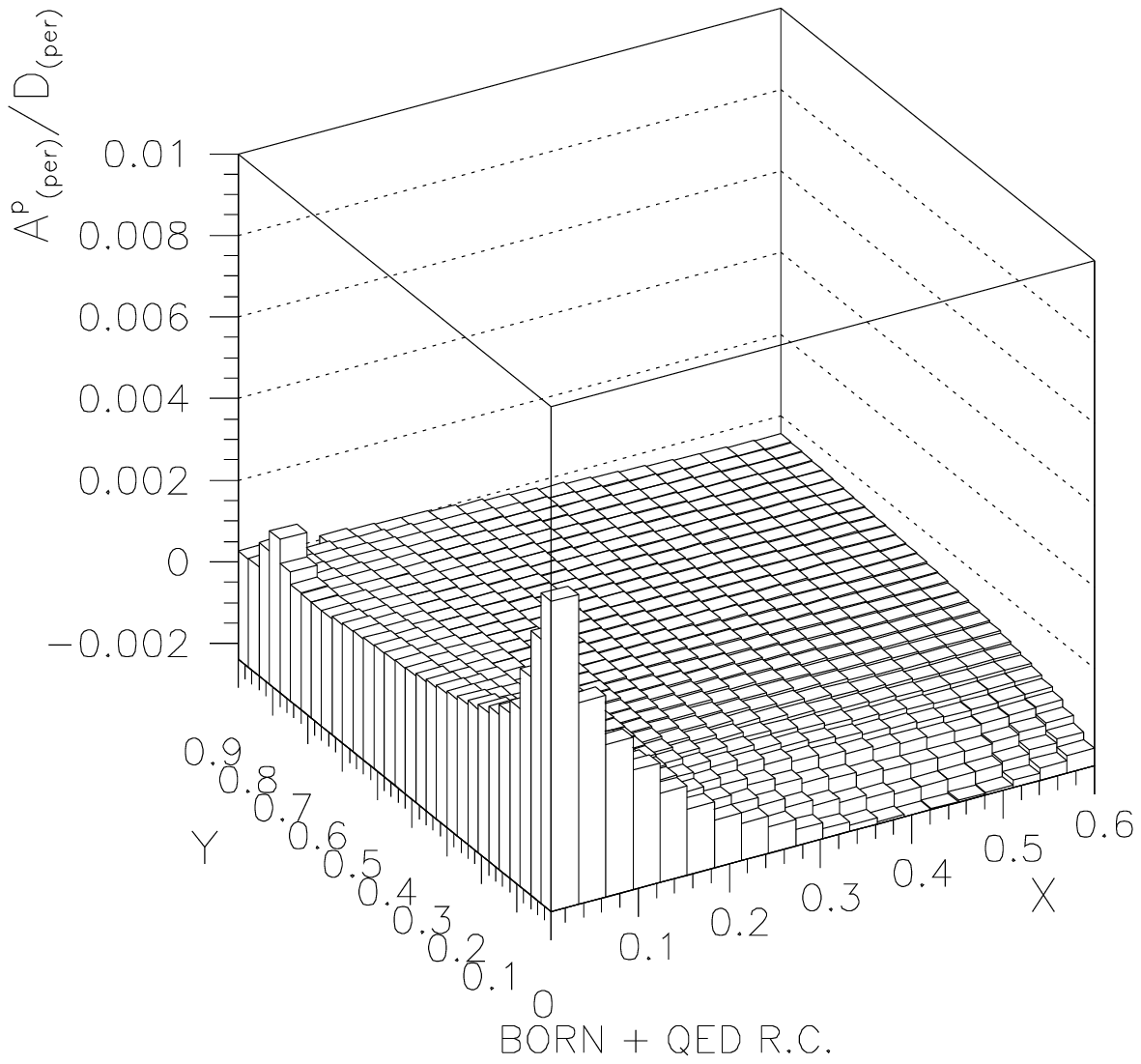}
\begin{center}
{Fig. 8.     QED corrected
               transverse  asymmetry $ A_{\perp}^p/
                                             D_{\perp}$
                          for beam energy $ 100~GeV^2$;
$g_1^p$  from ~\cite{Alt}.}
\end{center}
\end{figure}

\begin{figure}
\epsfysize22cm
\epsffile{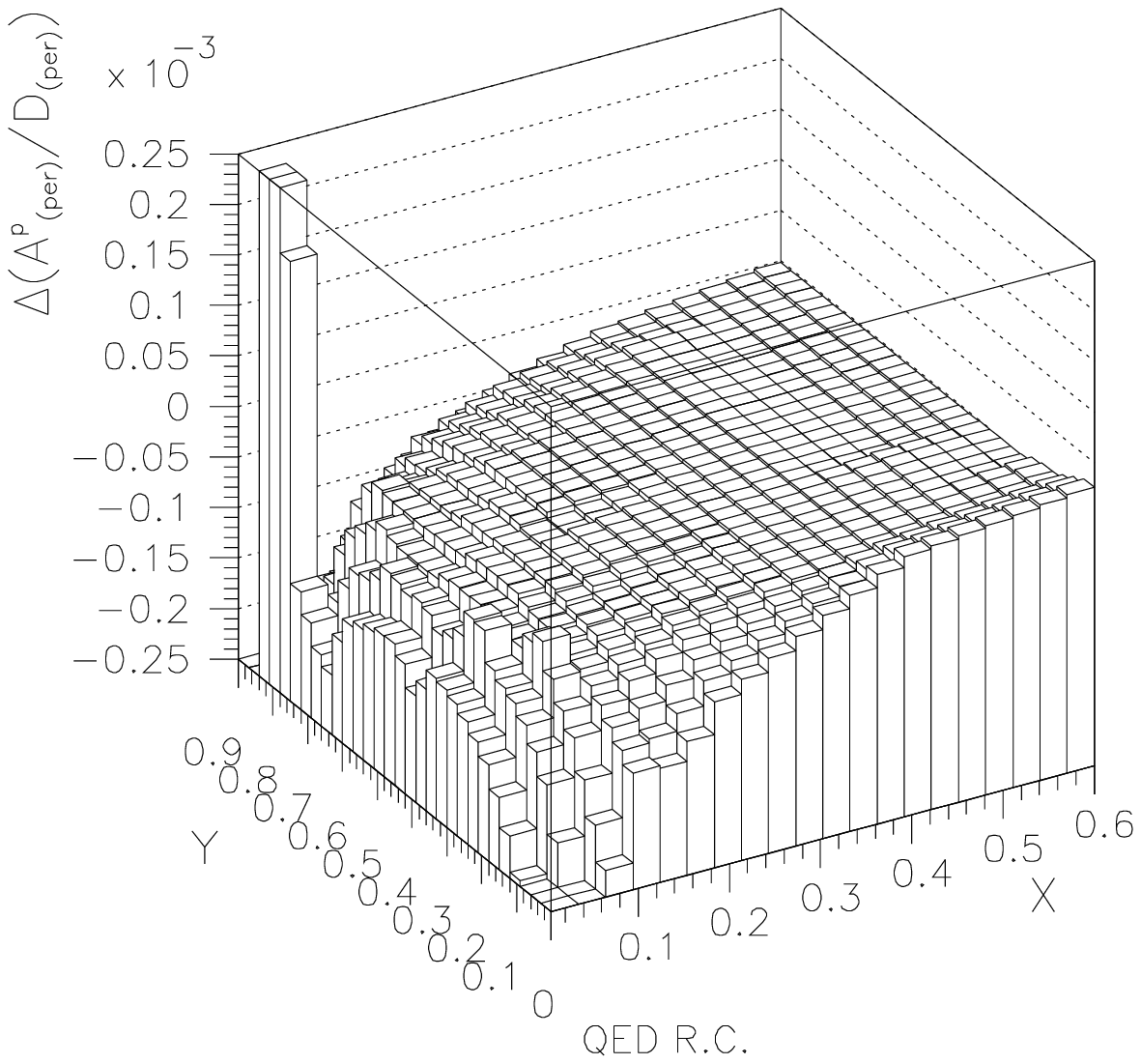}
\begin{center}
{Fig. 9.   Absolute correction  $ \Delta A_{\perp}^p/
                                             D_{\perp}$
                          for beam energy $ 100~GeV^2$;
$g_1^p$  from ~\cite{Alt}.}
\end{center}
\end{figure}
 
 
\end{document}

%% file: feinman.tex
\def\photonatomright{\begin{picture}(3,1.5)(0,0)
                                \put(0,-0.75){\tencircw \symbol{2}}
                                \put(1.5,-0.75){\tencircw \symbol{1}}
                                \put(1.5,0.75){\tencircw \symbol{3}}
                                \put(3,0.75){\tencircw \symbol{0}}
                      \end{picture}
                     }

\def\photonatomup{\begin{picture}(1.5,3)(0,0)
                             \put(-0.75,3){\tencircw \symbol{3}}
                             \put(-0.75,1.5){\tencircw \symbol{2}}
                             \put(0.75,1.5){\tencircw \symbol{0}}
                             \put(0.75,0){\tencircw \symbol{1}}
                   \end{picture}
                  }

\def\photonright{\begin{picture}(30,1.5)(0,0)
                     \multiput(0,0)(3,0){10}{\photonatomright}
                  \end{picture}
                 }

\def\photonrighthalf{\begin{picture}(30,1.5)(0,0)
                     \multiput(0,0)(3,0){5}{\photonatomright}
                  \end{picture}
                 }

\def\photonup{\begin{picture}(1.5,30)(0,0)
                  \multiput(0,0)(0,3){10}{\photonatomup}
               \end{picture}
              }

\def\fermionurr{\begin{picture}(30,15)(0,0)
                        \put(-30,-15){\vector(2,1){15}}
                        \put(-15,-7.5){\line(2,1){15}}
                  \end{picture}
                 }

\def\fermiondrr{\begin{picture}(30,15)(0,0)
                        \put(0,0){\vector(2,-1){15}}
                        \put(15,-7.5){\line(2,-1){15}}
                  \end{picture}
                 }

\newenvironment{Feynman}[3]{\begin{center}
                            \setlength{\unitlength}{#3 mm}
                            \begin{picture}(#1)(#2)
                            \thicklines
                           }{\end{picture} \end{center}}